\documentstyle[12pt]{article}

\def\d{\delta}\def\l{\lambda}\def\L{\Lambda}\def\S{\Sigma}
\def\intchg{\leftrightarrow}

\begin{document}

\begin{center}
\large{{\bf Gravitational Constraints which Generate\\
a Lie Algebra}}
\end{center}

\vspace{20pt}
\begin{center}
by\\
Karel V. Kucha\v{r} and Joseph D. Romano\\
Department of Physics\\
University of Utah\\
Salt Lake City, UT 84112
\end{center}

\section*{Abstract}

The coupling of gravity to dust helps to discover simple quadratic
combinations of the gravitational super-Hamiltonian and supermomentum
whose Poisson brackets strongly vanish.
This leads to a new form of vacuum constraints which generate a true
Lie algebra.
We show that the coupling of gravity to a massless scalar field leads
to yet another set of constraints with the same property, albeit not
as simple as that based on the coupling to dust.

\bigskip
\noindent PACS number(s): 0460, 0420

\section*{1. Commuting variables based on dust}

By coupling geometry to dust, Brown and Kucha\v{r}\cite{BrownKuchar}
found simple quadratic combinations
\begin{equation}
G(x):=(H^G(x))^2-g^{ab}(x)H_a^G(x)H_b^G(x)
\label{eq:G}
\end{equation}
of the gravitational super-Hamiltonian $H^G(x)$ and supermomentum $H_a^G(x)$
which have strongly vanishing Poisson brackets among themselves:
\begin{equation}
\{G(x),G(x')\}=0\,.
\label{eq:GG}
\end{equation}
This allowed them to replace the constraint system
\begin{equation}
H^G(x)=0=H_a^G(x)
\label{eq:oldconstraints}
\end{equation}
of {\it vacuum} gravity by an equivalent system
\begin{equation}
G(x)=0=H_a^G(x)
\label{eq:newconstraints}
\end{equation}
which generates a true Lie algebra.
Indeed, because the gravitational supermomentum $H_a^G(x)$ represents
LDiff$\S$,
\begin{equation}
\{H_a^G(x),H_b^G(x')\}=H_b^G(x)\d_{,a}(x,x')-(ax\intchg bx')\,,
\label{eq:HGaHGb}
\end{equation}
and $G(x)$ transforms as a scalar density of weight +2 under Diff$\S$,
\begin{equation}
\{G(x),H_a^G(x')\}=G_{,a}(x)\d(x,x')+2G(x)\d_{,a}(x,x')\,,
\label{eq:GHGa}
\end{equation}
the new constraints (\ref{eq:newconstraints}) close according to the
Lie algebra (\ref{eq:GG}), (\ref{eq:HGaHGb}), and (\ref{eq:GHGa}).
The underlying group is the semidirect product of the Abelian group
generated by $G(x)$ with Diff$\S$.

In the derivation of Eq.~(\ref{eq:G}), dust played the role of a catalyst.
When gravity is coupled to dust, $G(x)$ equals the square of the density
$P(x)$ of the rest mass measured by the Eulerian observers whose worldlines
are orthogonal to a hypersurface $\S$.
Because $P(x)$ is the canonical momentum conjugate to the proper time
$T(x)$ along the dust worldlines,
\begin{equation}
\{P(x),P(x')\}=0\,,
\end{equation}
and the closing relation (\ref{eq:GG}) follows by a simple argument.

Equation (\ref{eq:GG}), however, is clearly an identity involving only the
geometric variables $g_{ab}(x)$ and $p^{ab}(x)$.
Thus, it must hold irrespective of whether geometry is coupled to dust,
to any other matter system, or whether it is left alone in vacuum.
Indeed, one can verify that Eq.~(\ref{eq:GG}) holds by virtue of the
Dirac ``algebra'' (see Eqs.~(\ref{eq:HH})--(\ref{eq:HaHb})) among
the gravitational super-Hamiltonian $H^G(x)$ and supermomentum $H_a^G(x)$.
In the end, the coupling to dust which led us to expression (\ref{eq:G})
for $G(x)$ can be forgotten.

It is natural to ask whether the coupling of gravity to other sources yields
alternative combinations of the gravitational super-Hamiltonian and
supermomentum whose Poissons brackets also strongly vanish.
In this paper, we illustrate how this can be done for the simple example of
a massless scalar field.

\section*{2. Commuting variables based on a massless scalar field}

A massless scalar field described by the canonical variables
$\phi(x)$, $\pi(x)$ has the energy density
\begin{eqnarray}
H^\phi(x)&=&T(x)+V(x)\nonumber\\
&=&{1\over 2}g^{-{1\over 2}}(x)\pi^2(x)+{1\over 2}g^{1\over 2}(x)
g^{ab}(x)\phi_{,a}(x)\phi_{,b}(x)\label{eq:Hphi}
\end{eqnarray}
and the momentum density
\begin{equation}
H_a^\phi(x)=\pi(x)\phi_{,a}(x)\,.
\label{eq:Hphia}
\end{equation}
It is coupled to gravity by adding these terms to the gravitational
super-Hamiltonian and supermomentum.
This leads to the constraints
\begin{equation}
H(x):=H^G(x)+H^\phi(x)=0
\label{eq:Htot}
\end{equation}
and
\begin{equation}
H_a(x):=H_a^G(x)+H_a^\phi(x)=0
\label{eq:Htota}
\end{equation}
on the phase space $(g_{ab}(x),p^{ab}(x),\phi(x),\pi(x))$ of the coupled
system.

Equations (\ref{eq:Hphi})--(\ref{eq:Htota}) enable us to express the scalar
field momentum $\pi(x)$ entirely in terms of the geometric variables.
First, Eq.~(\ref{eq:Hphia}) gives the product of the kinetic and potential
energy densities:
\begin{equation}
{1\over 4}g^{ab}H_a^\phi H_b^\phi=T\, V\,.
\label{eq:TV}
\end{equation}
Since the sum $T+V$ is given by Eq.~(\ref{eq:Hphi}), we get
\begin{equation}
|T-V|=\sqrt{(H^\phi)^2-g^{ab}H_a^\phi H_b^\phi}\,.
\label{eq:T-V}
\end{equation}
{}From (\ref{eq:Hphi}) and (\ref{eq:T-V}) we obtain $T$ and $V$ as
\begin{eqnarray}
T&=&{1\over 2}\left(H^\phi\pm\sqrt{(H^\phi)^2-g^{ab}H_a^\phi H_b^\phi}
\,\right)\,,\label{eq:T}\\
V&=&{1\over 2}\left(H^\phi\mp\sqrt{(H^\phi)^2-g^{ab}H_a^\phi H_b^\phi}
\,\right)\,,\label{eq:V}
\end{eqnarray}
where the upper sign holds for $T>V$ and the lower sign for $T<V$.
{}From $H^\phi$ and $g^{ab}H_a^\phi H_b^\phi$ alone we cannot decide which
choice of sign is correct because Eqs.~(\ref{eq:Hphi}) and (\ref{eq:TV})
are symmetric under the interchange of $T$ with $V$.

By using the constraints (\ref{eq:Htot}) and (\ref{eq:Htota}), we can
express $T$ and hence $\pi^2$ as a functional of the geometric data:
\begin{equation}
\pi^2(x)=\L_{\pm}(x;g,p]:=g^{1\over 2}(x)\left(-H^G(x)\pm\sqrt{G(x)}
\,\right)\,,
\label{eq:Lpm}
\end{equation}
where $G(x)$ is given by Eq.~(\ref{eq:G}).

Depending on what choice of sign is correct, the constraints
(\ref{eq:Htot}) and (\ref{eq:Htota}) imply the constraints
\begin{equation}
C(x):=\pi^2(x)-\L(x;g,p]=0\,,
\label{eq:C}
\end{equation}
where $\L$ is either $\L_+$ or $\L_-$.
Because the constraints (\ref{eq:Htot}) and (\ref{eq:Htota}) are first
class, the constraints (\ref{eq:C}) must also be first class; i.e.,
\begin{equation}
\{C(x),C(x')\}=\{\L(x),\L(x')\}\approx 0\,,
\label{eq:CC}
\end{equation}
where the weak equality $\approx$ stands for ``modulo the constraints
(\ref{eq:Htot}) and (\ref{eq:Htota}).''
Because the bracket $\{\L(x),\L(x')\}$ depends only on the geometric
variables $g_{ab}(x)$ and $p^{ab}(x)$, while the constraints (\ref{eq:Htot})
and (\ref{eq:Htota}) contain also the scalar field variables $\phi(x)$
and $\pi(x)$, one may try to argue that (\ref{eq:Htot}) and (\ref{eq:Htota})
cannot help to turn the bracket (\ref{eq:CC}) into zero, so that it actually
must strongly vanish:
\begin{equation}
\{\L(x),\L(x')\}=0\,.
\label{eq:LL}
\end{equation}

Unfortunately, unlike the corresponding argument for incoherent
dust\cite{BrownKuchar}, the present argument has a loophole.
True, the constraints (\ref{eq:Htot}) and (\ref{eq:Htota}) contain the
scalar field variables, but certain combinations of them do not.
In particular, Eq.~(\ref{eq:Hphia}) tells us that $H_a^\phi$ is
proportional to a gradient and hence
\begin{equation}
\d^{abc}H_a^\phi H_{b,c}^\phi = 0\,.
\end{equation}
The momentum constraint (\ref{eq:Htota}) thus implies the constraint
\begin{equation}
\d^{abc}H_a^G H_{b,c}^G = 0
\label{eq:hiddenconstraint}
\end{equation}
which involves only the geometric data.
The bracket (\ref{eq:LL}) could thus turn out to vanish only weakly,
modulo the constraint (\ref{eq:hiddenconstraint}).

Inspite of this loophole in our reasoning, the conjecture (\ref{eq:LL})
is actually true.
We shall prove it by brute force in the Appendix.
This calculation has nothing to do with how we originally arrived at the
expression (\ref{eq:Lpm}) for $\L$.
As in the case of dust, Eq.~(\ref{eq:LL}) holds irrespective of whether
geometry is coupled to a scalar field, to any other matter system, or
whether it is left alone in vacuum.
Also it holds for both $\L_+$ and $\L_-$.

As before, we can replace the original set (\ref{eq:oldconstraints})
of vacuum constraints by a new set
\begin{equation}
\L(x)=0=H_a^G(x)
\end{equation}
which generates a true Lie algebra.

The new expressions $\L_\pm(x)$ contain as a building block the quadratic
combination $G(x)$ which naturally arises from coupling gravity to dust.
It is far from trivial, however, that the expressions $\L_\pm(x)$ have
strongly vanishing Poisson brackets, as $G(x)$ did.
The cross terms among $g^{1\over 2}$, $H^G$, and $G$ do not vanish, and
they cancel only if these constituents are combined exactly as in
Eq.~(\ref{eq:Lpm}).
In particular, let us note that the weight +1 densities
\begin{equation}
\l_\pm:=-H^G\pm \sqrt G
\end{equation}
{\it do not} have strongly vanishing Poisson brackets.

Note also that the square root $G^{1\over 2}$ in Eq.~(\ref{eq:Lpm}) leads to
terms containing $G^{-{1\over 2}}$ in the Poisson bracket
(\ref{eq:LL}).
These are individually ill-defined on the constraint surface.
For $G\not=0$, all such terms cancel and Eq.~(\ref{eq:LL}) follows.
It is thus important to stay away from the constraint surface when proving
Eq.~(\ref{eq:LL}), and pass to it only at the end.

Other couplings may lead to yet another combination of the gravitational
super-Hamiltonian and supermomentum with strongly vanishing Poisson brackets.
Brown found some such combinations when studying couplings to perfect
fluids\cite{Brown}.
It remains to be investigated if the multiplicity of such alternative
commuting constraints conveys some general message about the structure of
canonical gravity.

\section*{Acknowledgments}

Discussions with David Brown during the initial stages of this work are
gratefully acknowledged.
This research was supported in part by the NSF grants PHY89-04035
and PHY-9207225, and by the U.S.-Czech Science and Technology Grant
No~92067.

\section*{Appendix: Calculation of the Poisson Brackets}

In this appendix, we show by direct calculation that the Poisson bracket
of $\L_\pm(x)$ with $\L_\pm(x')$ is strongly equal to zero.
The goal is to give just enough detail to make the calculation transparent,
yet not to provide so much detail as to bore the reader.

Because all calculations in this part involve only the gravitational
variables, we shall omit the superscript $G$ on the constraints.
Thus, $H^G$ and $H_a^G$ are now called $H$ and $H_a$.
Let us begin by breaking $\L_\pm(x)$ into simpler pieces:
\begin{eqnarray}
\L_\pm(x)&:=&g^{{1\over 2}}(x)\l_\pm(x)\,,\label{eq:Lambda}\\
\l_\pm(x)&:=&-H(x)\pm\sqrt{G(x)}\,,\label{eq:lambda}\\
G(x)&:=&(H(x))^2 - F(x)\,,\label{eq:Gapp}\\
F(x)&:=&g^{ab}(x)H_a(x) H_b(x)\,,\label{eq:F}
\end{eqnarray}
where
\begin{eqnarray}
H(x)&:=&G_{abcd}(x)p^{ab}(x)p^{cd}(x)-g^{1/2}(x) R(x),\label{eq:HG}\\
H_a(x)&:=&-2g_{ab}(x) D_c p^{bc}(x)\,,\label{eq:HGa}
\end{eqnarray}
are the gravitational super-Hamiltonian and supermomentum, and
\begin{equation}
G_{abcd}(x):={1\over 2}g^{-{1\over 2}}(x)\bigg(g_{ac}(x)g_{bd}(x)
+g_{ad}(x)g_{bc}(x)-g_{ab}(x)g_{cd}(x)\bigg)
\label{eq:Gabcd}
\end{equation}
is the gravitational supermetric.
$R(x)$ is the scalar curvature and $D_{a}$ is the covariant derivative
associated with the metric $g_{ab}(x)$.
To simplify the notation further, we will henceforth omit the $\pm$
subscripts, and we will not bother to write spatial arguments
$x$ or $x'$.
Instead, we will denote the spatial argument of a field by putting a
prime (or no prime) on the stem letter of the field.
For example, $H'$ will serve as a shorthand notation for $H(x')$,
$H'_a$ for $H_a(x')$, $\L$ for $\L_\pm(x)$, etc.

Our calculation will use the facts that $H$ and $H_a$ have Poisson brackets
satisfying the Dirac algebra\cite{Dirac}, and that the Poisson brackets of
$F$ with $F'$ and $G$ with $G'$ strongly vanish\cite{BrownKuchar}.
We therefore take as given the fundamental Poisson brackets
\begin{equation}
\{g_{ab},p'{}^{cd}\}={1\over 2}(\d_a^c\d_b^d + \d_b^c\d_a^d)\d(x,x')\,,
\end{equation}
the Dirac algebra\cite{Dirac} Poisson brackets
\begin{eqnarray}
\{H,H'\}&=&g^{ab} H_a\d_{,b}(x,x') - (x\intchg x')\,,\label{eq:HH}\\
\{H,H'_a\}&=&H_{,a}\d(x,x')+H\d_{,a}(x,x')\,,\label{eq:HHa}\\
\{H_a,H_b'\}&=&H_b\d_{,a}(x,x') -(ax\intchg bx')\,,\label{eq:HaHb}
\end{eqnarray}
and the previously calculated\cite{BrownKuchar} Poisson brackets
\begin{equation}
\{F,F'\}=0=\{G,G'\}\,.
\label{eq:FFGG}
\end{equation}

In the above expressions, $\d(x,x')$ should be thought of as a quantity
that transforms as a scalar function in the first argument $x$ and as a
scalar density of weight +1 in the second argument $x'$.
An important identity involving the $\d$-function is
\begin{equation}
f(x')\d_{,a}(x,x')=f(x)\d_{,a}(x,x')+f_{,a}(x)\d(x,x')
\label{eq:identity}
\end{equation}
where $f(x)$ is a scalar function.
In fact, the identity (\ref{eq:identity}) still holds if $f(x)$ is
replaced by any tensor field with density weight 0.
The identity (\ref{eq:identity}) enables us to change the arguments of the
coefficients of $\d_{,a}(x,x')$.
The additional terms proportional to $\d(x,x')$ cancel in calculations
involving the interchange $(x\intchg x')$.
We will make use of this fact in the following calculations.
\vskip .2cm

We now list a series of six short steps which, when taken together, yield
the desired result: $\{\L,\L'\}=0$.
\vskip .2cm

{\bf Step 1}: Show that
\begin{equation}
\{\L,\L'\}=g^{1\over 2}g'{}^{1\over 2}\,\{\l,\l'\}+\left(g'{}^{1\over 2}
\l\,\{g^{1\over 2},\l'\}-(x\intchg x')\right)\,.
\label{eq:step1}
\end{equation}
\noindent The proof follows from a repeated use of the Leibniz rule
$\{f,gh\}=\{f,g\}h+g\{f,h\}$ applied to the product functions
$\L=g^{1\over 2}\l$ and $\L'=g'{}^{1\over 2}\l'$.
\vskip .2cm

{\bf Step 2}: Show that
\begin{equation}
\{g^{1\over 2},\l'\}=\mp g^{1\over 2}G^{-{1\over 2}}g^{ab}H_a\d_{,b}(x,x')+
\,\propto \, \d(x,x')\,,
\label{eq:step2}
\end{equation}
\noindent where $\propto\, \d(x,x')$ stands for ``terms proportional to
$\d(x,x')$.''
To verify Eq.~(\ref{eq:step2}), express $\l'$ in terms of $H'$
and $H'_a$ using Eqs.~(\ref{eq:lambda})--(\ref{eq:F}) and note
that $\{g{}^{1\over 2},H'\}\propto\, \d(x,x')$ since $H'$ is an ultralocal
function of the momentum $p'{}^{ab}$.
Thus, obtain
\begin{eqnarray}
\{g^{1\over 2},\l'\}&=&\mp{1\over 2}G'{}^{-{1\over 2}}\{g^{1\over 2},F'\}
+\,\propto\, \d(x,x')\nonumber\\
&=&\mp G'{}^{-{1\over 2}} g'{}^{ab}H'_a\{g^{1\over 2},H'_b\}
+\,\propto\, \d(x,x')\,.
\label{eq:step2a}
\end{eqnarray}
Since $g^{1\over 2}$ transforms as a scalar density of weight +1 under
Diff$\S$,
\begin{equation}
\{g^{1\over 2},H'_b\}=g^{1\over 2}\d_{,b}(x,x')
+\,\propto\, \d(x,x')\,.
\label{eq:step2b}
\end{equation}
Equations (\ref{eq:step2a}) and (\ref{eq:step2b}) together with the
identity (\ref{eq:identity}) yield (\ref{eq:step2}).
\vskip .2cm

{\bf Step 3}: Show that
\begin{equation}
\left(g'{}^{1\over 2}\l\,\{g^{1\over 2},\l'\}- (x\intchg x')\,\right)=
\mp g^{1\over 2}g'{}^{1\over 2}G^{-{1\over 2}}\l g^{ab}H_a\d_{,b}(x,x') -
(x\intchg x')\,.
\label{eq:step3}
\end{equation}
\noindent This step follows directly from step 2 since the terms proportional
to $\d(x,x')$ vanish when we antisymmetrize with respect to $x$ and $x'$.
Steps 1, 2, and 3 represent the first half of the Poisson bracket calculation.
\vskip .2cm

{\bf Step 4}: Show that
\begin{equation}
2HH'\{H,H'\}=H'\{F,H'\}-(x\intchg x')\,.
\label{eq:step4}
\end{equation}
\noindent This step follows as an immediate consequence of (\ref{eq:FFGG})
and the Leibniz rule for products.
Simply expand the RHS of $0=\{G,G'\}$ using (\ref{eq:Gapp}) and
(\ref{eq:FFGG}) to obtain the above result.
\vskip .2cm

{\bf Step 5}: Show that
\begin{equation}
\{\l,\l'\}=\{H,H'\}\mp\left({1\over 2}G^{-{1\over 2}}HH'{}^{-1}\{H,F'\}-
(x\intchg x')\,\right)\,.
\label{eq:step5}
\end{equation}
\noindent To obtain this result, use (\ref{eq:lambda}) to express $\l$ in
terms of $H$ and $G$.
Then use (\ref{eq:Gapp}), (\ref{eq:FFGG}), and the result of step 4
to simplify the Poisson brackets on the RHS.
\vskip .2cm

{\bf Step 6}: Show that
\begin{equation}
g^{1\over 2}g'{}^{1\over 2}\{\l,\l'\}=\pm g^{1\over 2}g'{}^{1\over 2}
G^{-{1\over 2}}\l g^{ab}H_a\d_{,b}(x,x')-(x\intchg x')\,.
\label{eq:step6}
\end{equation}
\noindent This step follows from step 5 once we express $F'$ in terms of
$g'{}^{ab}$ and $H'_a$ via (\ref{eq:F}) and use the Dirac algebra
relations (\ref{eq:HH}) and (\ref{eq:HHa}) to evaluate the Poisson brackets
$\{H,H'\}$ and $\{H,H'_b\}$.
The identity (\ref{eq:identity}) must also be used to eliminate the primes
on some of the quantities.
Steps 4, 5, and 6 represent the second half of the Poisson bracket
calculation.
\vskip .2cm

By inspecting steps 1, 3, and 6, we conclude that
\begin{equation}
\{\L(x),\L(x')\}=0\,.
\end{equation}
This is the desired result (\ref{eq:LL}).

\end{document}